# Secure Edge Computing Reference Architecture for Data-driven Structural Health Monitoring: Lessons Learned from Implementation and Benchmarking

Sheikh Muhammad Farjad
University of Nebraska at Omaha
Omaha, Nebraska, USA
sfarjad@unomaha.edu

Sandeep Reddy Patllola
University of Nebraska at Omaha
Omaha, Nebraska, USA
spatllola@unomaha.edu

Yonas Kassa
Robert Morris University
Pittsburgh, Pennsylvania, USA
ykassa@acm.org

George Grispos
University of Nebraska at Omaha
Omaha, Nebraska, USA
ggrispos@unomaha.edu

Robin Gandhi
University of Nebraska at Omaha
Omaha, Nebraska, USA
rgandhi@unomaha.edu

## Abstract

Structural Health Monitoring (SHM) plays a crucial role in maintaining aging and critical infrastructure, supporting applications such as smart cities and digital twinning. These applications demand machine learning models capable of processing large volumes of real-time sensor data at the network edge. However, existing approaches often neglect the challenges of deploying machine learning models at the edge or are constrained by vendor-specific platforms. This paper introduces a scalable and secure edge-computing reference architecture tailored for data-driven SHM. We share practical insights from deploying this architecture at the Memorial Bridge in New Hampshire, US, referred to as the Living Bridge project. Our solution integrates a commercial data acquisition system with off-the-shelf hardware running an open-source edge-computing platform, remotely managed and scaled through cloud services. To support the development of data-driven SHM systems, we propose a resource consumption benchmarking framework called edgeOps to evaluate the performance of machine learning models on edge devices. We study this framework by collecting resource utilization data for machine learning models typically used in SHM applications on two different edge computing hardware platforms. edgeOps was specifically studied on off-the-shelf Linux and ARM-based edge devices. Our findings demonstrate the impact of platform and model selection on system performance, providing actionable guidance for edge-based SHM system design.

## CCS Concepts

• **Computer systems organization** → **Embedded and cyber-physical systems**; *Real-time systems*; • **Computing methodologies** → **Machine learning**.

## Keywords

edge computing, structural health monitoring, machine learning

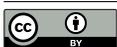

## 1 Introduction

Aging transportation infrastructure elements pose a critical safety concern. Recent incidents, such as bridge collapses [9, 24], highlight the urgent need for advanced data-driven monitoring of infrastructures to enable timely action and prevent loss of lives and assets. The move towards smart cities and the incorporation of aging infrastructures into such systems further emphasizes the importance of robust structural health monitoring systems.

Advances in sensor technologies have made possible various structural characteristics to be monitored. These include strain [12], temperature, displacement [21], tilt [14], corrosion [29], vibration [15], and structural defects such as cracks [13]. However, the use of individual sensors in silos has inherent limitations in terms of providing a comprehensive understanding of the health of a monitored system of interconnected structural components. Notably, diverse elements like bridge decks (concrete, gravel, timber, etc.), superstructures and substructures require specialized sensors to capture their unique characteristics, as they have distinct wear and tear patterns [19]. Integrating multiple sensors and data streams offers a promising direction for a comprehensive understanding of structural health.

By combining data from various sensors, a structural health monitoring (SHM) system can enable a more accurate and robust assessment of infrastructure health. For instance, integrating image-based crack detection with displacement sensor data, as demonstrated in Zaurin et al. [32], can provide a more comprehensive analysis and prediction of potential failures. Furthermore, such integrated availability of SHM data and analytics can enable innovative data-driven applications beyond currently manual inspection, maintenance and repair processes. For instance, route planning and optimization can be enhanced by considering the real-time structural health of bridges and roads, ensuring safer and more efficient transportation systems [20]. However, SHM approaches based on batch data collection and processing cannot scale or provide timely information to stakeholders. The complexities of multi-sensor heterogeneous data, varying sampling rates, and advanced analytics requirements now demand robust, scalable, and secure edge computing resources with machine learning capabilities that can be provisioned and managed centrally. Furthermore, supporting advanced applications for structural health monitoring such as digital twinning demands large volumes of real-time sensor data streams to be collected and integrated at the network's edge. However, existing approaches often neglect the challenges of data fusion or deploying machine learning models at the edge. Many commercial solutions are also limited to vendor-specific platforms. This paper introduces a scalable edge-computing reference architecture tailored for data-driven SHM. We share practical insights from deploying this architecture at the Memorial Bridge in New Hampshire, US, also known as the Living Bridge project [30]. Our solution integrates a commercial data acquisition system with off-the-shelf hardware running an open-source edge-computing platform, remotely managed and scaled through cloud services. To support the development of data-driven SHM systems, we also propose a benchmarking framework to evaluate the performance of machine learning (ML) inference operations on edge devices.

In summary, main contributions of this paper are:

- A reference architecture for supporting scalable and secure structural health monitoring with edge devices, with broad applicability to multiple domains.
- A case study of implementing the proposed reference architecture for SHM at the Portsmouth Memorial Bridge in New Hampshire, US, which hosts the Living Bridge project [30].
- A benchmarking framework designed to evaluate the resource utilization of ML inference operations on Linux and ARM based edge devices. Using this framework, we perform a comprehensive analysis of resource utilization metrics — including CPU usage, inference latency, and memory consumption — for machine learning models typically used in SHM applications. A detailed evaluation is presented for a Convolutional Neural Network (CNN) model, while the implementation of the framework for other models is briefly discussed. The experiments are carried out on off-the-shelf edge devices, namely the Raspberry Pi 4 and BeagleBone AI-64.

The remaining sections are organized as follows. In Section 2, we present existing architectural paradigms in SHM literature discussing their benefits and limitations. In Section 3, we present a scalable edge-computing reference architecture tailored for data-driven SHM. In Section 4, we share practical insights from implementing this architecture as part of an existing SHM system. We then present a benchmarking framework for edge devices in Section 5. In Section 6, we discuss experimental results and provide implementation recommendations for future work. Finally, Section 7 concludes the paper.

## 2 Literature Review

SHM is an extensively researched field with multiple approaches aimed at improving the safety and reliability of civil infrastructures, such as buildings, bridges, and railways [22, 23]. Mishra et al. [23] provide a comprehensive review of SHM systems leveraging Internet of Things (IoT) devices, detailing case studies for high-rise buildings, railways, and bridges. However, while many studies focus on individual SHM components, few provide an end-to-end solution that integrates all aspects of SHM, including data acquisition, processing, and monitoring.

For example, Gatti [16] presented a case study of monitoring an operational bridge, covering data acquisition with sensors, but did not provide a comprehensive framework for orchestrating the entire monitoring process. Similarly, Xu et al. [31] developed a testbed for damage simulation in bridges, yet their solution is limited to laboratory settings and relies on commercial software, limiting its adoption for more diverse applications. Additionally, Al Harrasi et al. [2] proposed a cybersecurity-focused testbed to evaluate vulnerabilities in SHM components, emphasizing the importance of securing wireless components. Aguzzi et al. [11] highlighted the susceptibility of SHM machine learning models to adversarial attacks, illustrating a critical challenge in ensuring the security and robustness of SHM platforms.

Several studies have also explored edge computing architectures for SHM applications. Buckley et al. [8] introduced an Edge-SHM framework using low-power wireless sensing, demonstrating how long-range, low-power IoT-driven edge platforms can facilitate continuous monitoring. Their system integrates microcontrollers with local computation of damage-sensitive features before transmission to a cloud platform. Similarly, Hidalgo-Fort et al. [18] proposed a low-cost, low-power edge computing system for bridge SHM, utilizing modular hardware with onboard processing capabilities to minimize data transmission over NB-IoT networks. Their architecture integrates synchronized time-series data analysis and local modal identification, enhancing SHM efficiency.

The use of machine learning at the edge in SHM has been gaining attention, offering enhanced damage detection and predictive maintenance capabilities [6]. Oliveira et al. [26] explored the integration of embedded systems, edge computing, and TinyML in the Internet of Intelligent Things (IoIT) paradigm, emphasizing autonomous decision-making on resource-constrained devices. Similarly, Hao et al. [17] examined deep learning frameworks on edge devices, demonstrating that inference performance is heavily influenced by hardware constraints, batch sizes, and framework choices. While these studies highlight the potential of ML at the edge in SHM applications, efficient deployment remains a challenge due to the limited computational resources of edge devices. Azimi et al. [6] reviewed various statistical and ML-based damage detection methods that can be integrated into edge-based SHM systems, yet they do not address the critical issue of resource consumption (memory, CPU, and inference time) when deployed on edge hardware. For practical deployments, optimizing resource utilization can be as important a consideration as ML predictive accuracy and explainability.

Although significant progress has been made in SHM research, a gap remains in the literature regarding the deployment of efficient, secure, and scalable SHM systems. Our work contributes to filling this gap by proposing a secure edge computing reference architecture for data-driven SHM, along with a benchmarking tool for evaluating the resource utilization of machine learning models on edge computing devices.

## 3 Edge Computing Reference Architecture for Data-driven Structural Health Monitoring

Our proposed reference architecture promotes loose coupling among system components through a layered design. A layered design allows existing components from different vendor platforms to be easily accommodated or replaced in an integrated system. In our experience, a three-layered design comprising perception, edge, and cloud layers provides enough information domains where context-relevant computing and network resources can be available with appropriate security controls. Each layer allows the segmentation and localization of faults, improving the security and maintainability of the system. For example, granular access control policies can be defined and provisioned within each layer. Figure 1 provides a high-level overview of our three-layered reference architecture. The following sections are used to describe each layer.

### 3.1 Perception Layer

The perception layer includes different types of sensor platforms, wired or wireless local area networks and data aggregators, which can receive real-time data from multiple channels at high sampling rates. This layer typically starts with the sensors and terminates with data aggregators. For SHM applications, authentication or encryption of raw sensor data can be implemented at terminal nodes that power the sensors to provide end-to-end data security. Terminal nodes connect with data aggregators through wired or wireless local area networks. Data aggregators, also referred to as data loggers, can collect multi-modal sensor data from multiple terminal nodes and package them for further processing. The data modality varies according to the requirements of the monitored structure and its environment. Contact-based SHM typically collects raw data from strain transducers, accelerometers and seismometers [28]. Similarly, vision-based SHM involves recording images to detect collisions or defects such as cracks, spalling, or corrosion and estimating the deterioration of the structural elements [13].

Data aggregators process sensor data streams by serializing them at configurable time intervals. This process consolidates the data into disk-savable files, enriched with metadata to enable efficient information discovery and retrieval. Depending on the type of data, data aggregators use various standard file formats. For instance, in contact-based SHM applications, Technical Data Management Solution (TDMS) files are used for storing and transferring numerical sensor data along with metadata [25]. TDMS is a hierarchical file format that follows three levels of hierarchy: file, group, and channel. Each file contains multiple groups, and each group consists of different channels, as illustrated in Figure 2. Different properties are assigned to each level of the hierarchy in TDMS files.

### 3.2 Edge Layer

The edge layer supports intelligence deployment closer to the data available from the perception layer. It can also reduce the amount of data that needs to be transmitted to the cloud layer, while supporting autonomous operations and buffer data during times of network unavailability. For data-driven SHM applications, an edge layer runtime environment enables scalable and remote deployment of machine learning models for local inference tasks and software updates.

In an ideal deployment, data aggregators in the perception layer should transmit sensor data to the edge layer using cross-domain access control policies. These policies should enforce one-way updates and incorporate file integrity checks. The edge layer devices process the data files received from the perception layer. Additionally, edge devices interface with cloud resources to offload computations or store data as needed.

An edge runtime is a lightweight software framework designed to operate on edge devices allowing for localized processing. It acts as an intermediary layer between the hardware of the edge device and the applications or models deployed on it, facilitating efficient execution and communication along with portability. For example, the edge device runtime enables the deployment of machine learning models that are trained in the cloud using large datasets to perform inference tasks locally. Inference tasks generally require significantly less computing resources compared to model training and optimization tasks. Depending on the edge device's resources (such as CPU, memory, and network bandwidth) and the size of data files, one of the following deployment strategies can be used: **(a) Local inference:** If the device has sufficient resources (CPU and memory) to process data efficiently, the inference component can run directly on the device; or **(b) Cloud-based inference:** If the device lacks adequate computational power but has reliable network bandwidth, the data can be transmitted to inference components hosted in the cloud for further processing. Local inference deployment strategies are preferred for on-site inspections and diagnostics of real-time structural response to loads independent of cloud layer availability. The edge runtime can also support distributed processing of compute-intensive tasks for next-generation UAV-based visual inspections.

### 3.3 Cloud Layer

The cloud layer offers essential services such as data storage, remote management of the edge runtime, and advanced analytics for building SHM applications. With access to large SHM datasets, training and optimization of machine learning models can be performed in the cloud, which can then be deployed at edge devices for inference tasks. Cloud-native services, when paired with a compatible edge runtime, enable the remote deployment of software and machine learning models to edge devices. This capability is particularly valuable for edge devices located in remote areas with limited physical access.

Edge devices are designed to operate efficiently in rugged environments, prioritizing power efficiency and accommodating limited storage capacities. As a result, cloud storage services periodically fetch data from the edge devices for longitudinal data collection.

### 3.4 Cross-layer Communication

To facilitate secure and reliable data exchange, layers communicate with their neighboring layers through various modes of communication. The perception layer interacts with the edge layer *via* sensor data APIs, which are developed based on protocols selected according to specific design decisions or services available for vendor platforms [7]. Similarly, communication between the edge layer and the cloud layer relies on deployment and data exchange APIs.

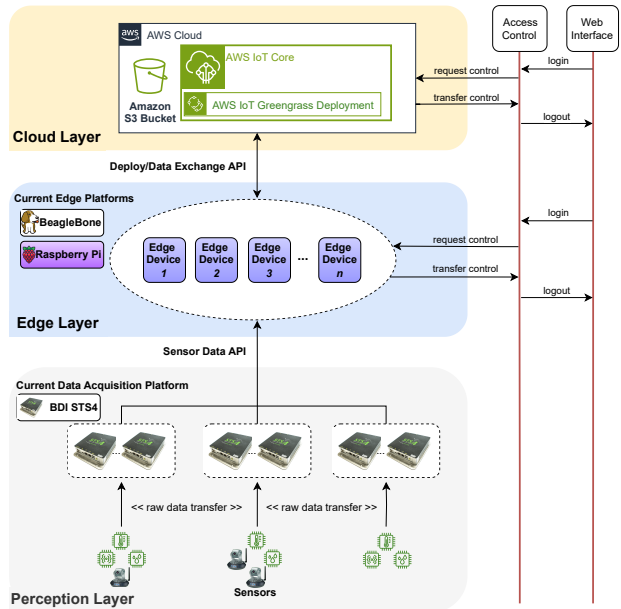

Figure 1: 3-tier Architecture of SHM Testbed

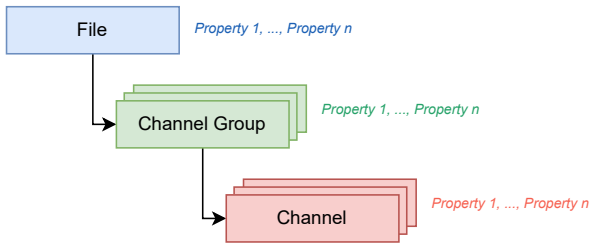

Figure 2: Hierarchical Structure of TDMS File

These APIs differ from sensor data APIs as they perform two primary tasks: remote deployment and data exchange. Developers can use remote deployment APIs to update deployments on an edge device runtime by redeploying specific software components. Similarly, data exchange APIs enable authenticated and encrypted data transfers between edge devices and the cloud environment.

Cross-layer communication APIs ensure that messages are authenticated and encrypted to mitigate the risk of eavesdropping attacks. Beyond communication security, the overall security of each layer depends on the measures implemented by the components within that specific layer. The segmentation introduced by the layered architecture ensures that security controls are tailored to the specific requirements of each layer based on risk management practices. Notably, perception layers and edge layers must also be secured against physical tampering to ensure system integrity.

## 4 Implementating the Reference Architecture

The reference architecture described in the previous section was informed and validated by implementing it at an existing SHM deployment. Our field deployment site is the World War I Memorial Bridge along the U.S. Route 1. It is a vertical-lift bridge across the Piscataqua River between New Hampshire and Maine. As a living laboratory, also referred to as the Living Bridge project [30], the bridge is instrumented with sensors that continuously monitor its structural elements and are powered by a tidal turbine. We describe the implementation of our three layered architecture, integrating it with the Living Bridge project SHM instrumentation. Various components of this implementation as shown in Figure 1 are further described in this section.

### 4.1 Perception Layer

In the perception layer, we had access to strain gauge sensor streams from the vertical guide post frame members of the bridge. These sensors communicate the raw data streams to a single data aggregator housed in the bridge control cabin. The data aggregator is a STS4 Core Data Logger [7] device. It is a ruggedized industrial data acquisition system developed by BDI Inc. for operation in harsh environments, making it particularly suitable for SHM applications. Equipped with preconfigured software components, the STS4 Logger allows for the configuration of the data sampling rate from connected sensors as well as serializing the data into TDMS files at certain time intervals. The data logger includes an optional component, called the data bridge, which transfers the TDMS files to a designated FTP server. This data bridge service exports data from the perception layer to devices in the edge layer using a one-way FTP protocol.

## 4.2 Edge Layer

Our implementation uses a Raspberry Pi 4 in the edge layer. This edge device hosts two primary services, one for interfacing with the perception layer and the other for communicating with the cloud layer. To receive TDMS files from the perception layer, the edge device run an FTP service. The Raspberry Pi also hosts an open-source AWS IoT Greengrass edge runtime and cloud service. It enables edge devices to process and analyze data locally, run ML models, respond autonomously to events, and securely communicate with other edge devices and AWS services. Using AWS IoT Greengrass, we have deployed a software component to analyze and compress the data in TDMS files for sensor streams by computing a Root Mean Square (RMS) value for each sensor stream. The computed RMS value is reported to the cloud layer. Due to limited storage on Raspberry Pi, we have also implemented a background systemd daemon to periodically transfer all TDMS files to AWS cloud storage using AWS IoT Greengrass cloud services. This ensures seamless data transfer from the sensor node to the cloud through an authenticated channel established by the services running on the Raspberry Pi.

## 4.3 Cloud Layer

To enable remote deployment of software components and ML models to the edge runtime along with cloud integration, our implementation uses AWS IoT Greengrass [5]. Once processed edge data is available in the cloud in S3 object storage, we used AWS QuickSight [4] as a cloud-hosted dashboard. Figure 3 shows a tile in the QuickSight dashboard that visualizes the RMS values for the top and bottom sensor channels for Members 3 and 7 computed by the edge runtime and collected in the cloud over a few months. Within QuickSight, we have also enabled anomaly detection and alert generation features to monitor any significant deviation of RMS values from their baseline values. Figure 4 shows one such alert in the QuickSight dashboard.

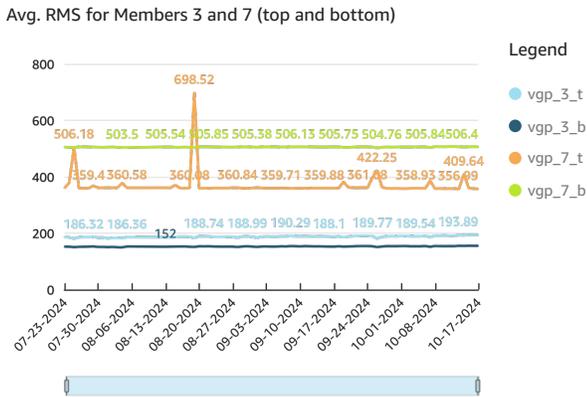

Figure 3: RMS Values from AWS QuickSight Dashboard

## 4.4 Cross-layer Communication

For communication between the STS4 Data Logger and the Raspberry Pi, the reference implementation relies on FTP data transfer over a wired connection, as both devices are housed within a physically safe and locked cabin. STS4 Data Logger currently does not support SFTP. However, the use of an Ethernet interface for FTP communication within a physically secure area minimizes cybersecurity risks.

The Raspberry Pi edge runtime uses X.509 certificates for authentication with cloud services and relies on AWS IoT policies for authorization within AWS IoT Greengrass. X.509 digital certificates adhere to the X.509 public key infrastructure (PKI) standard, which associates a public key with the identity specified in the certificate. Meanwhile, IoT policies enforce precise authorization controls for components interacting with different segments of AWS IoT Greengrass operations, such as data plane manipulations [3].

**Anomaly detection**

LAST UPDATED AN HOUR AGO

The most recent anomaly detected on Oct 14, 2024 11am was:
- Average vgp_7_t was 512.18, which was **higher** than the expected 426.72

Figure 4: Anomaly Detection in AWS QuickSight Dashboard

# 5 Benchmarking ML Models on Edge Devices

To support the development of data-driven SHM systems, we now describe a benchmarking framework to understand the resource utilization of inference tasks using machine learning models on edge devices. Using this framework, we compare the resource utilization of machine learning models typically used in SHM applications, deployed on off-the-shelf Linux and ARM-based edge devices. Next, we describe our instrumentation design and its use for benchmarking ML model inference operations on resource-constrained edge devices.

## 5.1 Instrumentation - *edgeOps*

To understand and measure the performance of ML inference operations on edge devices, we developed a tool called edgeOps[1]. The primary goal of edgeOps is to capture the resource utilization of inference models with a minimal memory or computational burden on the edge devices. edgeOps is currently capable of capturing memory utilization, CPU consumption, and latency of inference operations.

*Memory Profiler.* This component is responsible for calculating the memory usage of model inference tasks on edge devices. It relies on decorators from a Python memory profiler library, which are directly embedded in the inference code of ML programs (i.e., inference.py in Figure 5). The inference.py file also contains additional blocks of code which can be used for feature extraction before model inference.

*Latency Module.* To calculate the execution time at a more granular level, edgeOps utilizes the built-in date utility with nanosecond precision in a Linux environment. This module is designed with consideration for potential failures that might occur during the execution of inference components. During development, we observed that the execution of ML inference must be synchronized with the `date` utility to accurately capture inference time. Without synchronization, the `date` utility may record erroneous values due to a speed mismatch with ML inference. To implement this safe synchronization, we introduce padding chunks of 30 seconds before and after the execution of the inference component. These additional chunks are discarded when calculating the execution time. This time padding ensures that the model has fully recovered from the previous execution cycle. As shown in Figure 5, the main block of `edgeOps.sh` incorporates this functionality directly.

---
[1]Our *edgeOps* implementation code is available to the research community at: https://github.com/smfarjad/edgeOps

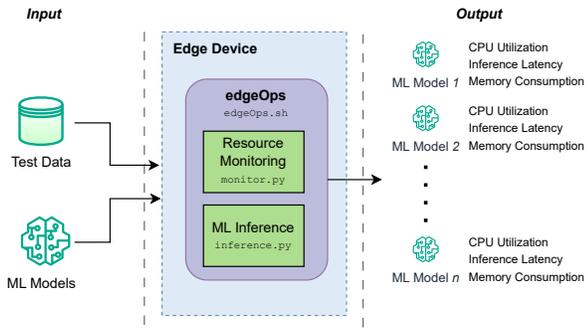

Figure 5: Architecture of `edgeOps`

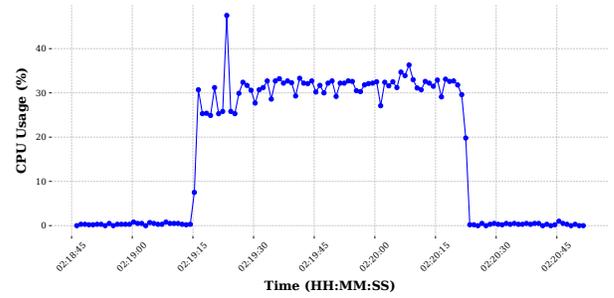

Figure 6: Plot for CPU Usage

*CPU Usage Component.* This component is responsible for capturing the average CPU usage during the model inference of an ML classifier. The `edgeOps` framework utilizes the `psutil` library to calculate the CPU usage of the device during inference. Since edge devices often have background processes that consume CPU resources, we developed a strategy to accurately reflect the CPU usage attributable to the inference component. To achieve this, `edgeOps` calculates the CPU usage of the edge device within two defined time windows: (a) when the inference component is running on the edge device, and (b) when the device is not running any inference. For the time window (a), we record the CPU usage while the code is executing and compute an average value (indicated by elevated peaks starting at the timestamp 02:19:15 in Figure 6). For time window (b), we utilize CPU usage recorded during the padded 30-second timing windows inserted before and after inference execution to obtain the average CPU usage during idle operation. After calculating the average CPU usage for both time windows, we determine the difference, which represents the additional CPU usage resulting due to execution of the inference component.

The `monitor.py` script implements this functionality of calculating CPU usage and generates $n$ CSV files, where $n$ represents the number of iterations `edgeOps` executes the ML model with its corresponding test data. Algorithm 1 illustrates the entire process of measuring resource consumption using `edgeOps`. Note that `edgeOps` conducts measurements over a batch of 100 inputs repeated 10 times. This technique allows a larger time duration for measurements and filters out noise-related variations using mean values.

**Algorithm 1:** Resource Consumption Measurement

$M \leftarrow$ get_trained_models() // get trained models to be loaded to devices
$B \leftarrow 100$ // batch size
$DD \leftarrow$ [Device1, Device2] // edge devices
$R \leftarrow 10$ // number of repetitions
**foreach** *device* $\in DD$ **do**
  // perform experiment on device
  **foreach** *model* $\in M$ **do**
    $ML \leftarrow$ port_model_to_device(**model**, **device**)
    **for** $i \leftarrow 1$ **to** $R$ **do**
      restart_machine() // optional
      load_model_into_memory(**device**, **ML**)
      start_monitoring(cpu, timestamp)
      $M \leftarrow [\,]^B$ // memory usage
      **for** $j \leftarrow 1$ **to** $B$ **do**
        start_record_memory_consumption()
        $I \leftarrow$ load_image(j)
        $P \leftarrow$ perform_prediction(**device**, **I**)
        $M[j] \leftarrow$ record_memory_consumption()
      **end**
      stop_monitoring(cpu, timestamp)
    **end**
    aggregate_measurements(**cpu**, **timestamp**, **M**)
  **end**
  report_results()
**end**

## 5.2 Evaluation

In this section, we describe the experimental setup developed to evaluate the operability of `edgeOps` by measuring the inference performance of a Convolutional Neural Network (CNN) based classifier for an SHM application.

*Data Acquisition.* Data acquisition in SHM relies on three main approaches: contact-based methods, non-contact-based methods, and vision-based methods [27]. Contact-based methods involve the installation of sensors directly on the structure, which can be challenging to install and may affect the structural properties and responses. Although non-contact methods use wireless sensors, their installation and operation can also be complex and expensive, often requiring highly skilled workers. These limitations make vision-based methods more favorable, as they offer high accuracy, a non-contact approach, and lower costs. Furthermore, we argue that vision-based approaches are particularly suitable for the edge computing paradigm, especially in scenarios where edge devices are expected to generate results in real time. For instance, a drone equipped with a vision-based model can inspect structures and generate results in real time without relying on Internet connectivity. Such scenarios are particularly advantageous in contested environments where data security is important and adversaries may attempt to compromise the data. These considerations form the basis for our decision to assess a vision-based classifier based on CNN using edgeOps.

*Dataset.* We use a surface crack detection dataset to train a CNN-based model. The dataset [34] consists of images of concrete surfaces divided into two categories: negative (without cracks) and positive (with cracks). The images are organized into separate folders to streamline the classification process. Each class includes 20,000 images, resulting in a total of 40,000 images. All images have dimensions of 227 × 227 pixels and are in RGB format. The dataset was created from 458 high-resolution images (4032 × 3024 pixels) using the method described by Zhang et al. [33]. These source images exhibit variations in surface texture and lighting conditions. No data augmentation techniques, such as rotation or flipping, were applied to the dataset.

*Training.* We split the crack detection dataset into training and testing sets using an 80-20 ratio, enabling the model to be evaluated on unseen data. For the CNN, the TensorFlow library [1] was used to develop a simple CNN classifier for crack detection. Parameters were adjusted during the development phase to ensure the algorithm operated effectively on the crack detection dataset. Since the objective was not to design an efficient or novel classifier, we did not focus on optimizing the algorithm to achieve specific performance thresholds. We trained the CNN model on a Linux server equipped with an Intel(R) Xeon(R) Gold 5317 CPU @ 3.00 GHz. Our CNN model consists of three convolutional layers (with 32, 64, and 128 filters respectively) followed by max-pooling layers to progressively reduce the spatial dimensions of the feature maps. After the convolutional blocks, the model flattens the feature maps and passes them through two fully connected layers, with 512 neurons and a final output layer for binary classification. A dropout layer is included to prevent overfitting. The model has a total of 19,034,177 trainable parameters, optimizing its performance for complex image recognition tasks. In our evaluation, the trained CNN model achieved an F1 Score of 99.2. Since the F1 Score is the harmonic mean of Precision and Recall, it provides a balanced evaluation by considering both metrics simultaneously [10].

*Edge Platforms for Deployment.* After training, the model was deployed on two edge devices selected for evaluation. To measure the performance of ML inference operations, a set of 100 test images was also transferred to the edge platforms. Our experimental setup consists of two system-on-chip (SoC) platforms: the TDA4VM and BCM2711, which are used in the popular IoT platforms BeagleBone AI-64 and Raspberry Pi 4 respectively. These SoC platforms provide sufficient computational resources for IoT applications, supporting complex operations such as image processing and the execution of microservices. Table 1 presents the system specifications for each platform.

*Using* edgeOps. The first step to using edgeOps is to prepare the inference code (i.e., **inference.py**) and import the decorator for collecting memory usage above the relevant block of code. In our case, we created a function that takes the path to the input image, preprocesses the image, performs prediction, and returns the results. Since edgeOps relies upon memory_profiler, we decorate the corresponding function with @profile as shown in the following code.

```
@profile
def calling_decorators(image_path):
    img = preprocess_image(image_path)
    prediction = model.predict(img)
    return prediction
```

The output of **inference.py** is redirected to the file **edgeOps.sh**, which allows for persistent storage of output data, enabling later review and analysis.

```
local start_time=$(date +%s.%N)
python3 inference.py >> mem_record.txt
local end_time=$(date +%s.%N)
```

The variables start_time and end_time are used to measure the inference latency of the ML program (i.e., **inference.py**). These timestamps are embedded in the filenames of the CSV files generated by **monitor.py**, which log CPU usage along with their corresponding timestamps using the psutil library. Each iteration generates a separate CSV file. With the current configuration of edgeOps set to 10 iterations, our experiment using the CNN model produces one text file and ten CSV files.

## 6 Results and Discussion

After analyzing the data obtained from edgeOps, we present the following results highlighting resource utilization of our CNN-based classifier on both edge platforms (i.e., Raspberry Pi 4 and BeagleBone AI-64), as shown in Figure 7.

Figure 7(a) presents the *CPU usage* of the two devices. The Raspberry Pi demonstrated significantly lower mean CPU usage, averaging 29.96%, compared to 53.02% for the BeagleBone AI-64. The error bars, indicating the standard deviation, reveal minimal variability in CPU usage across multiple runs for both devices. This observation demonstrates the efficiency of the 4-core Raspberry Pi in managing computational tasks with reduced overall CPU usage, compared to a 2-core Beaglebone AI-64 device.

Figure 7(b) presents the inference *latency results*. Over the 100 images, the Raspberry Pi achieved an average latency of 67.73 seconds, which is slightly lower than the BeagleBone AI-64's average of 67.56 seconds. The error bars indicate minor variations in latency for both devices, demonstrating comparable stability in real-time inference performance. Note that despite a higher clock speed, a 2-core platform does not have a significant advantage in latency over a 4-core platform with a slower clock speed.

Figure 7(c) focuses on *memory utilization*, where the Raspberry Pi again outperformed the BeagleBone AI-64 in terms of resource efficiency. The average memory usage for the Raspberry Pi was 548.44 MB, while the BeagleBone AI-64 consumed 691.43 MB on average. The error bars suggest a slightly higher variability in memory usage for the BeagleBone AI-64, possibly reflecting differences in memory allocation strategies or system overheads. These results highlight the trade-offs between resource utilization and performance across the two edge devices. The Raspberry Pi consistently demonstrated better resource efficiency, with lower CPU and memory usage, while the BeagleBone AI-64 consumed more resources but showed slightly more stable inference times.

Table 1: Platform Specifications

| SoC Platform | Processor | Architecture | # of cores | Clock speed | Memory | GPU |
|---|---|---|---|---|---|---|
| TDA4VM | ARM Cortex-A72 | 64-bit | 2 | 2.0GHz | 4GB | PowerVR Rogue 8XE GE8430 |
| BCM2711 | ARM Cortex-A72 | 64-bit | 4 | 1.5GHz | 4GB | Videocore VI |

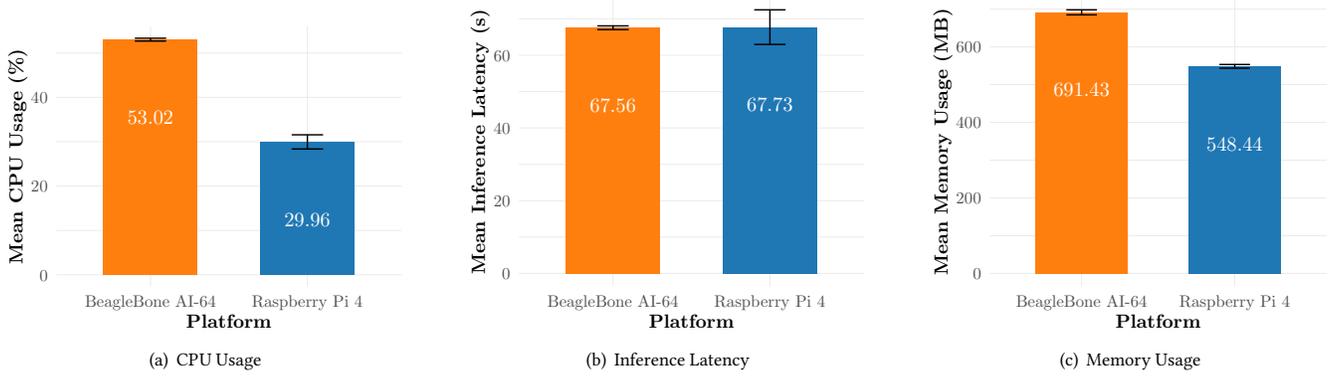

(a) CPU Usage  (b) Inference Latency  (c) Memory Usage

Figure 7: Performance Metrics of CNN Algorithm on Raspberry Pi 4 and BeagleBone AI-64

Figure 8 offers an alternative comparison of these findings. Each edge device is represented by a triangle, with the vertices corresponding to inference latency (runtime), CPU usage, and memory consumption. A smaller triangular area indicates better resource efficiency, as it reflects lower values across all three metrics, except for runtime, where performance is similar on both platforms. These insights are valuable for optimizing edge device selection according to specific application requirements. Our comparisons focus on inference tasks, which are inherently CPU-intensive. We have prioritized model portability over device-specific optimizations for these tasks, operating under the assumption that models are trained in the cloud and later deployed to resource-constrained edge devices. However, for applications that require training AI models directly on edge devices, platforms like the Beaglebone-AI 64, which offers better GPU capabilities, would be more suitable. In our experiments, we did not enable any device-specific model optimizations or utilize GPU libraries.

**Additional algorithms.** We adopted the same strategy using edgeOps, transitioning from training on a server to inference on edge devices, to evaluate classifier implementations based on Logistic Regression (LR), Support Vector Machine (SVM), and k-Nearest Neighbor (k-NN) algorithms. While SVM and LR demonstrated consistent performance, the k-NN classifier failed to operate under standard settings on either edge platform. To address memory constraints, memory overcommitment was enabled on the Raspberry Pi 4 and BeagleBone AI-64 by modifying /proc/sys/vm/overcommit

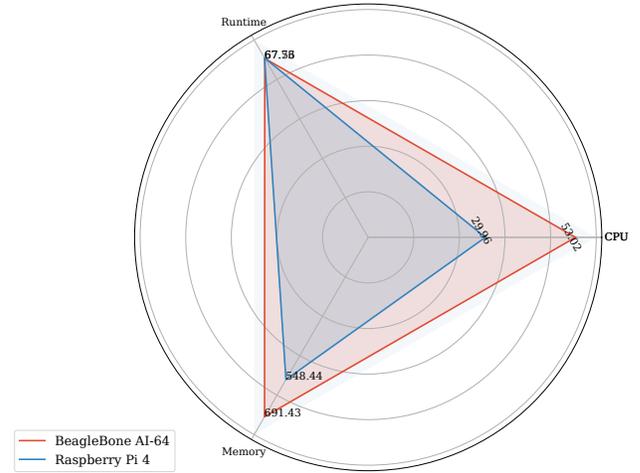

Figure 8: Radar Plot for Raspberry Pi 4 and Beaglebone AI-64

_memory, allowing for increased memory allocation during ML inference. Despite these modifications, k-NN did not run successfully on the BeagleBone AI-64 and exhibited high memory usage on the Raspberry Pi 4. This behavior can be attributed to the lazy learner characteristic of k-NN, which does not construct a discriminative model during training but instead retains the entire dataset and organizes it for efficient querying. As a result, k-NN showed deviations in performance when compared to SVM and LR, both of which build models during training.

To further assess the resource demands of k-NN, the dataset size was reduced from 40k to 5k images. Under these conditions, k-NN operated seamlessly on both devices, delivering consistent results and confirming compatibility with the edgeOps framework for reduced-scale datasets.

**Limitations.** The primary objective of edgeOps is to measure the resource requirements of ML models for inference tasks across various edge platforms. It is not designed to evaluate the performance or optimization of ML model training or the prediction quality of the models. Additionally, while ML inference can be optimized for the targeted hardware, it comes at the cost of model portability. For instance, in its default setting, the inference code does not utilize available GPUs on the platforms, which limits the exploitation of hardware capabilities. However, to maintain consistency and accurately capture the resource requirements of ML models for inference tasks across different edge platforms, we kept the inference code identical for both platforms in our study.

## 7 Concluding Remarks

In this paper, we presented a reference architecture for deploying data-driven SHM systems within an edge computing paradigm. Our architecture was validated in a real-world environment, demonstrating its capability for continuous and reliable monitoring. Furthermore, we introduced edgeOps, a benchmarking tool designed to evaluate the resource utilization of deployed ML models on edge devices. Using edgeOps, we presented results from assessing a CNN-based classifier on two edge device platforms, the BeagleBone AI-64 and Raspberry Pi 4. To ensure repeatability, we provided detailed instructions for using edgeOps, enabling ML practitioners and researchers to adapt it for their own models.

While our work offers a robust framework for resource-aware SHM systems at the edge, there are several avenues for future work. First, existing tools for energy measurement are predominantly designed for Intel–based architectures. Hence, new approaches are needed for energy computation monitoring in Linux and ARM–based edge devices. Second, the need for platform-specific model optimization can lead to variations between inference models, which have not addressed in this research. Finally, future work can also investigate extending our benchmarking capabilities to support other constraints, such as energy efficiency across diverse architectures, and integrating optimization strategies tailored to specific edge platforms.

## Acknowledgments

We thank the anonymous reviewers for their valuable feedback. This work is partially supported by contracts W912HZ21C0060 and W912HZ23C0005, US Army Engineering Research and Development Center (ERDC), and Award Number 1762034 from the National Science Foundation.

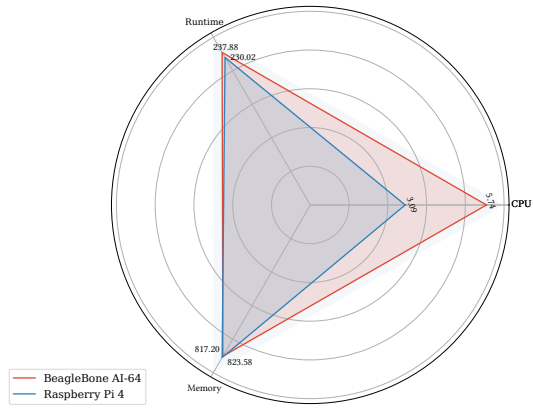

Figure 10: Radar Plot for SVM

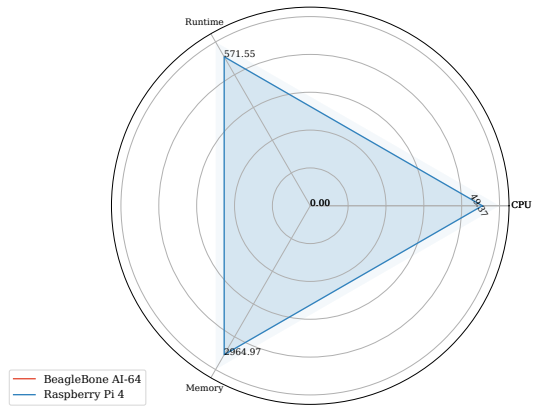

Figure 11: Radar Plot for k-NN

## A Appendix

### A.1 Supplemental Information

Figures 9, 10, and 11 display radar plots for the Logistic Regression (LR), Support Vector Machine (SVM), and k-Nearest Neighbors (k-NN) machine learning models, respectively. These plots are based on performance benchmarks of these models on Raspberry Pi 4 and BeagleBone AI-64 obtained using edgeOps. The k-NN model was not able to function on the BeagleBone AI-64. As a result the radar plot in Figure 11 does not include any data for this platform.

It is important to note that we did not optimize models or fine-tune their performance during our study. Additionally, the models vary significantly in their architectures and training methods. The purpose of our study was to assess whether our performance benchmarking methods could effectively highlight differences in model performance across different platforms. The radar plots presented illustrate these variations. However, since we did not optimize the models for specific use cases, these plots should not be used for direct comparisons between them. The primary goal of these visualizations is to demonstrate how edgeOps captures the resource utilization of machine learning algorithms across two distinct edge platforms.

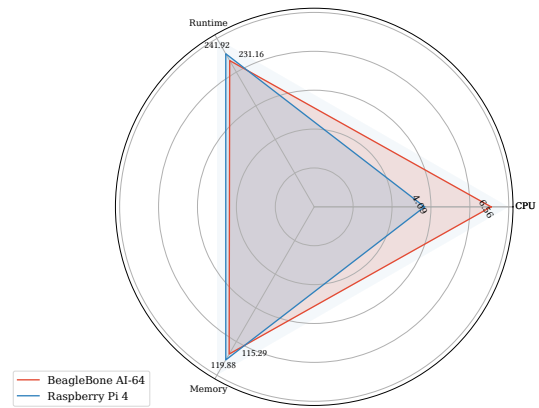

Figure 9: Radar Plot for LR